\documentclass[a4paper]{jpconf}
\usepackage{graphicx,color}
\usepackage{amssymb}
\usepackage{amsmath}
\definecolor{red}{rgb}{0.8,0,0}

\usepackage{gensymb}
\usepackage{amsmath,amssymb,epsfig,bm,mathrsfs,color}
\usepackage{slashed}
\usepackage{color}
\newcommand{\be}{\begin{equation}}
\newcommand{\ee}{\end{equation}}
\newcommand{\bea}{\begin{eqnarray}}
\newcommand{\eea}{\end{eqnarray}}

\newcommand{\veck}{{\bm k}}
\newcommand{\vecQ}{{\bm Q}}

\newcommand{\vecr}{\bm r}

\newcommand{\SD}{^3S_1$-$^3D_1}

\def\curl{\mathop{\rm curl}}

\definecolor{red}{rgb}{0.8,0,0}
\definecolor{orange}{rgb}{0.8,0.2,0.0}
\definecolor{blue}{rgb}{0.3,0.0,0.8}
\definecolor{green}{rgb}{0,0.5,0.0}
\definecolor{darkred}{rgb}{0.7,.1,.2}
\definecolor{bgred}{rgb}{1.,.95,.95}
\definecolor{bgblue}{rgb}{.95,.95,1.}
\definecolor{bluegreen}{rgb}{0.,.5,.3}
\definecolor{darkred}{rgb}{0.7,.1,.2}
\definecolor{darkgreen}{rgb}{0.1,.6,.0}
\definecolor{lightyellow}{rgb}{1.,1.,.8}
\definecolor{darkcyan}{rgb}{0.,.7,.9}
\definecolor{lightblue}{rgb}{0.6,0.8,1}
\definecolor{lightgreen}{rgb}{0.7,1.,.9}
\definecolor{money}{rgb}{0.4,0.8,0.}
\definecolor{purple}{rgb}{0.9,0.0,0.8}
\definecolor{orange}{rgb}{0.9,0.5,0.0}
\definecolor{newgr}{rgb}{0.2,0.8,0.2}
\definecolor{newbl}{rgb}{0.3,0.6,0.8}
\definecolor{newor}{rgb}{1.0,0.6,0.}
\def\prb{Phys. Rev. B \ }%
\def\prc{Phys. Rev. C\ }%
\def\prd{Phys. Rev. D \ }%
\def\prl{Phys. Rev. Lett. \ }%
\begin{document}

\title{Toward electrodynamics of unconventional phases of dilute
  nuclear matter}
\author{ Armen Sedrakian}
\address{Frankfurt Institute for Advanced Studies, D-60438  Frankfurt am Main, Germany}
\ead{sedrakian@fias.uni-frankfurt.de}

\author{John W. Clark}
\address{Department of Physics and McDonnell Center for the Space
Sciences, Washington University, St.~Louis, Missouri 63130, USA}
\address{
Centro de Investiga\c{c}\~{a}o em Matem\'{a}tica e Aplica\c{c}\~{o}es,
University of Madeira, 9020-105 Funchal, Madeira, Portugal
}
\ead{jwc@wuphys.wustl.edu}

\begin{abstract}
   The phase diagram of isospin-asymmetrical nuclear matter may feature
  a number of unconventional phases, which include the translationally
  and rotationally symmetric, but isospin-asymmetrical BCS condensate,
  the current-carrying Larkin-Ovchinnikov-Fulde-Ferrell (LOFF) phase,
  and the heterogeneous phase-separated phase. Because the Cooper
  pairs of the condensate carry a single unit of charge, these phases
  are charged superconductors and respond to electromagnetic gauge
  fields by either forming domains (type-I superconductivity) or
  quantum vortices (type-II superconductivity). We evaluate the
  Ginzburg-Landau (GL) parameter across the phase diagram and find
  that the unconventional phases of isospin-asymmetrical nuclear
  matter are good type-II superconductors and should form Abrikosov
  vortices with twice the quantum of magnetic flux.  We also find that
  the LOFF phase at the boundary of the transition to the type-I
  state, with the GL parameter being close to the critical value
  $1/\sqrt{2}$. 
\end{abstract}

\section{Introduction}
\vskip 0.3cm

Understanding superconducting phases of nuclear matter at
sub-saturation densities is a challenging many-body physics problem.
The attraction in the nuclear force is responsible for the formation
of nuclear clusters, as well as condensates of
Bardeen-Cooper-Schrieffer (BCS) type at low temperatures, with
significant implications for the physics of supernovae and neutron
stars.  It has been conjectured such systems can exhibit the
celebrated BCS-BEC transition~\cite{Nozieres:1985zz}, i.e., a
transformation of loosely-bound Cooper pairs at weak coupling to bound
dimers at strong coupling that a form a Bose-Einstein condensate
(BEC)~\cite{1993NuPhA.551...45A,Baldo:1995zz,Lombardo:2001ek,2006PhRvC..73c5803S,2009PhRvC..79c4304M,Huang:2010fk,2016PhRvC..94c4004I,2017PhRvC..96c4327R,Fan:2017ioh}.
However, these nuclear systems are isospin asymmetric, which means
that the isoscalar neutron-proton ($np$) pairing is disrupted by the
mismatch in the Fermi surfaces of protons and
neutrons~\cite{2000PhRvL..84..602S}.  

In a recent series of papers~\cite{Stein:2012wd,Stein:2014nla,Stein:2015bpa}, 
the combined effects of the BCS-BEC crossover and mismatched Fermi surfaces have
been studied in the phase diagram of nuclear and neutron matter, see
also the reviews~\cite{2014JPhCS.496a2008S,Clark_review16}.  The
coupled equations for the gap and the densities of the constituents (neutrons 
and protons) have been solved allowing for (i) the ordinary BCS state, 
(ii) its low-density asymptotic counterpart BEC state, and (iii) two
phases that emerge once there is finite isospin asymmetry, namely the 
current-carrying phase (LOFF phase) and the spatially separated (PS-BCS)
phase.  The strong-coupling regime features the phase-separated BEC
(PS-BEC) phase. Another important characteristic of these phases --
the coherence length $\xi$ -- was extracted microscopically from the
wave functions of the Cooper pairs, thus avoiding the standard definition 
which is valid only in the BCS limit.

The present contribution focuses on the electromagnetic response of
the phases indicated above, which arises from the fact that the proton
in an $np$ Cooper pair carries a unit of charge. For this purpose
we evaluate the penetration depth $\lambda$ of the magnetic field in
various regions of the phase diagram and deduce the Ginzburg-Landau (GL)
parameter $\kappa = \lambda/\xi$. We then use the GL criterion to
establish the type of the superconductivity realized in the phase
diagram. 

We note here that the type-II versus type-I superconductivity of the
$S$-wave proton condensate in the core of a neutron star has been
widely discussed in the literature, see for the recent
work~\cite{2004PhRvL..92o1102B,2005PhRvC..72e5801A,2005PhRvD..71h3003S,2008PhRvB..78b4510A,2015PhRvC..91c5805S,2017PhRvD..95k6016H}
and for a recent review Ref.~\cite{2017arXiv170910340H}. The
proton superfluid contributes at most 10-15 percent to the density of
the core, therefore it actual density is of the same order as that of
the neutron-proton condensate considered here. The main differences
arise through larger gaps in the present case, which lead to smaller
coherence lengths, and the charges that the condensate carry (a unit
of charge in the present case).

This contribution is structured as follows. In Sec.~\ref{sec:review} we 
give a brief review of the theory of asymmetrical nuclear matter and discuss 
the resulting phase diagram.  In Sec.~\ref{sec:edyn} we address the 
electrodynamics of the superconducting state; we evaluate the GL parameter 
using input from our previous microscopic studies along with a new 
computation of the penetration depth.  In Section~\ref{sec:conclusions} 
we collect our conclusions and consider related perspectives.

\section{Review of the phase diagram}
\label{sec:review}
\vskip 0.3cm

Superfluid nuclear matter can be described in terms of the 
Nambu-Gor'kov propagators, which from a matrix 
\bea
\label{props} 
i\left(\begin{array}{cc} G_{12}^{+} & F_{12}^{-}\\
    F_{12}^+ & G_{12}^{-}\end{array}\right) = \left(\begin{array}{cc}
    \langle T_\tau\psi_1\psi_2^+\rangle
    & \langle T_\tau\psi_1\psi_2\rangle \\
    \langle T_\tau\psi_1^+\psi_2^+\rangle & \langle T_\tau\psi_1^+\psi_2\rangle
\end{array}\right),
\eea 
where $x = (t,\vecr) $ denotes the continuous temporal-spatial
variable and Greek indices label discrete spin and isospin variables.
Each operator in Eq.~(\ref{props}) can be viewed as a bi-spinor, i.e.,
$\psi_{\alpha}=(\psi_{n\uparrow},\psi_{n\downarrow},\psi_{p\uparrow},\psi_{p\downarrow})^T,$ where the symbols $\uparrow, \downarrow$ label a particle's
spin and the indices $n,p$ label its isospin.

In energy-momentum space the elements of the matrix \eqref{props}
are given by 
\bea
G_{n/p}^{\pm} &=&
\frac{ik_{\nu}\pm\epsilon_{p/n}^{\mp}}{(ik_{\nu}-E^+_{\mp/\pm})(ik_{\nu}+E^-_{\pm/\mp})},\\
F_{np}^{\pm} &=&
\frac{-i\Delta}{(ik_{\nu}-E^+_{\pm})(ik_{\nu}+E^-_{\mp})},\\
F_{pn}^{\pm} &=&
\frac{i\Delta}{(ik_{\nu}-E^+_{\mp})(ik_{\nu}+E^-_{\pm})},
\eea
where the four branches of the quasiparticle 
spectrum are given by 
\be 
\label{eq:QP_spectra}
E_{r}^{a} = \sqrt{E_S^2+\Delta^2} + r\delta\mu +aE_A, 
\ee 
in which $a, r \in \{+,-\}$.   Here 
\bea
\label{eq:E_S}
E_S =\frac{Q^2/4+k^2}{2m^*}-\bar\mu, \quad 
E_A = \frac{\veck\cdot \vecQ}{2m^*} ,
\eea 
are the symmetrical and anti-symmetrical parts of the quasiparticle
spectrum,  $\bar\mu \equiv (\mu_n+\mu_p)/2$, 
$m^*$ is the effective mass of nucleon, and $\vecQ$ is the center-of-mass
momentum of the Cooper pairs. 

The anomalous self-energy (pairing-gap) evaluated in the standard
fashion to lowest order in the neutron-proton interaction 
$ V(\veck,\veck')$ is expressed as
\bea \label{eq:gap}
\Delta(\veck,\vecQ) &=&  \frac{1}{4\beta} \int\!\!\frac{d^3k'}{(2\pi)^3}\sum_{\nu}
V(\veck,\veck')               \nonumber\\
&& {\rm Im}  \Bigl[  
  F^+_{np} (k'_{\nu},\veck',\vecQ)
+F^-_{np} (k'_{\nu},\veck',\vecQ)  
-F^+_{pn} (k'_{\nu},\veck',\vecQ) 
-F^+_{pn} (k'_{\nu},\veck',\vecQ) 
\Bigr],
\eea
which yields, after partial-wave expansion and
computation of the Matsubara sums, 
\bea \label{eq:gap2}
\Delta_l(Q) = \frac{1}{4}\sum_{a,r,l'} \int\!\!\frac{d^3k'}{(2\pi)^3}
V_{l,l'}(k,k') 
\frac{\Delta_{l'}(k',Q)}{2\sqrt{E_{S}^2(k')+\Delta^2(k',Q)}}[1-2f(E^r_a)],
\eea
where $V_{l,l'}(k,k')$ is now the interaction in the $\SD$ partial wave,
$f(\omega)=1/[\exp{(\omega/T)}+1]$, and $\Delta^2 =\sum_l \Delta_l^2$.

The densities of neutrons and protons are obtained from the propagator
after phase-space integration and summation over the discrete
(spin-isospin) variables:
\bea\label{eq:densities}
\rho_{n/p} (\vecQ)&=&\frac{2}{\beta}\int\!\!\frac{d^3k}{(2\pi)^3}\sum_{\nu} 
G^+_{n/p}(k_{\nu},\veck,\vecQ)\nonumber\\
&=&2 \int\!\!\frac{d^3k}{(2\pi)^3}
\Biggl[
\frac{1}{2}
\left(1+\frac{E_S}{\sqrt{E_S^2+\Delta^2}}\right) f(E^+_{\mp})
+\frac{1}{2}             \left(1-\frac{E_S}{\sqrt{E_S^2+\Delta^2}}\right) f(-E^-_{\pm})
\Biggr].
\eea 
In writing the thermodynamic quantities we distinguish the spatially
homogeneous and inhomogeneous cases and label the superfluid (S) or 
unpaired (N) quantities correspondingly.  The free energies of the
homogeneous phases can then be simply written as 
\be \label{eq:free} F_S =
E_S-TS_S,\quad F_N = E_N-TS_N, \ee
where $E$ is the internal energy (statistical average of the system
Hamiltonian) and $S$ denotes the entropy.  The free energy of the
heterogeneous superfluid phase (i.e. the phase in which there is a
separation of the normal and superfluid phases) is given by 
\bea \label{eq:free_mixed}
\mathscr{F}(x,\alpha) = (1-x) F_S(\alpha = 0) 
+ x F_N(\alpha \neq 0), \,\, (Q = 0),\nonumber\\
\eea
where $x$ here denotes the filling fraction of the unpaired component
and 
\be \alpha = \frac{\rho_n-\rho_p}{\rho_n+\rho_p} 
\ee 
is the density asymmetry. The net density is given by $\rho =\rho_n+\rho_p $.

There are four possible states that can appear in the temperature-density 
phase diagram:
\bea\label{eq:phases}
\begin{array}{llll}
Q = 0,  &\Delta \neq 0, & x = 0,& \textrm{BCS phase,}\\
Q \neq 0, & \Delta \neq 0, &x = 0,&\textrm{LOFF phase,} \\
Q = 0,  &\Delta \neq 0, & x \neq 0,&\textrm{PS phase,}\\
 Q = 0, & \Delta = 0, & x = 1, &\textrm{unpaired phase.} \\
\end{array}
\eea
Clearly, the phase with lowest free energy at any given temperature 
and density corresponds to the ground state of the nuclear matter. 
\begin{figure}[tb]
\begin{center}
\includegraphics[width=8.5cm,height=7cm]{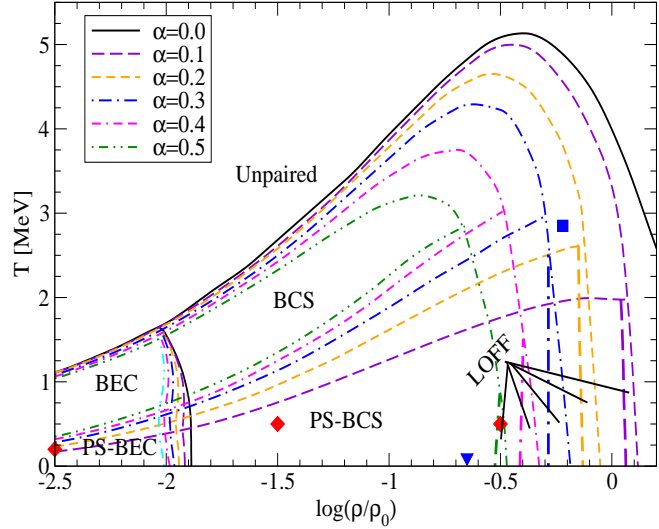}
\caption{Phase diagram of dilute nuclear matter in the temperature-density 
  plane for several isospin asymmetries $\alpha$ according to 
  Refs.~\cite{Stein:2012wd,Stein:2014nla}. The density is normalized 
  to the nuclear saturation density.  The phase diagram includes 
  the unpaired phase, the BCS (BEC) phase, the LOFF phase, and 
  the PS-BCS (PS-BEC) phase. For each asymmetry there are two 
  tri-critical points, one of which is always a Lifshitz point 
  \cite{Stein:2012wd,Stein:2014nla}. There is a single tetra-critical
  point in the phase diagram shown by the square dot.  Note that the
  LOFF phase exists in the lower right corner of the phase diagram. 
  The boundaries between BCS and BEC phases are identified by the 
  change of sign of the average chemical potential $\bar\mu$. 
  The diamonds mark the values of the density and temperature appearing
  in Table~\ref{tab:1}, these being chosen as representative of the 
  three coupling regimes.}

\label{fig:phasediagram}
\end{center}
\end{figure}
Their locations in the phase diagram are shown in
Fig.~\ref{fig:phasediagram} for several values of isospin asymmetry
$\alpha$.  One can identify four different phases classified according
to Eq.~\eqref{eq:phases}: (a) The unpaired phase occupies the 
high-temperature region $T>T_{c0}$ of the phase diagram, where
$T_{c0}(\rho)$ is the critical temperature of the normal/superfluid
phase transition at $\alpha =0$. (b) The current-carrying LOFF phase
is located in the lower right corner, corresponding to low
temperatures and high densities. (c) The PS phase occupies the lower
left corner, which corresponds to low densities and low temperatures.
As the temperature is increased, these last two phases transform into
(d) the isospin-asymmetrical BCS phase. 

To quantify the BCS-BEC crossover, we define three regimes of coupling
which we express in terms of densities.  The strong-coupling regime
(SCR) corresponds to the low-density limit where well-defined
deuterons exist. The weak-coupling regime (WCR) corresponds to the
high-density limit where well-defined Cooper pairs are formed.  The
regime intermediate between these domains is called the
intermediate-coupling regime (ICR).  Representative points in these
regimes are indicated by diamonds in Fig.~\ref{fig:phasediagram}.

\section{Evaluating the Ginzburg-Landau parameter}
\label{sec:edyn}
\vskip 0.3cm

BCS superconductors are characterized by at least three distinct
length scales: (i) the {\it London penetration depth} $\lambda$, (ii)
the {\it coherence length} $\xi$, and (iii) the mean interparticle distance
$d$.  The ratio of two of these scales defines the Ginzburg-Landau
(GL) parameter: $\kappa= \lambda/\xi$. In the range
$ 1/\sqrt{2} < \kappa <\infty,$ the material is a type-II
superconductor; otherwise it is type-I.  In conventional type-II
superconductors with Cooper charge $2e$ the magnetic field is carried
by electromagnetic vortices with quantum flux $\phi_0 = \pi\hbar c/e$.

\begin{figure}[tb]
\begin{center}
\includegraphics[width=8.5cm,height=7cm]{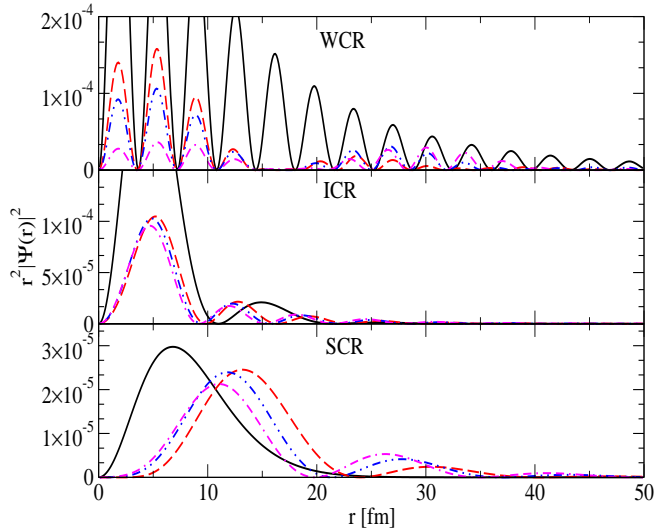}
\caption{ Dependence of $r^2|\Psi(r)|^2$ on $r$ for the three 
  coupling regimes~\cite{Stein:2014nla}.  (See Table \ref{tab:1} for corresponding
  values of density and temperature).  For the asymmetry parameter
  we consider values $\alpha = 0$ (black, solid), 0.1 (red, long-dashed), 0.2
  (blue, dash--double-dotted) and 0.3 (violet, double-dash-dotted).}
\label{fig:r_Psi}
\end{center}
\end{figure}

We now review  the evaluation the coherence length, which can be related 
to the root-mean-square radius of the Cooper-pair wave function, defined 
as \cite{Stein:2012wd,Stein:2014nla}
\be \label{eq:r2psi}
\langle r^2\rangle = \int d^3r\, r^2 \vert
\Psi(\vecr)\vert^2,  \ee 
where $\Psi(\vecr)$ is the Cooper-pair wave function.
Fig.~\ref{fig:r_Psi} shows the integrand of Eq.~\eqref{eq:r2psi},
i.e., the quantity $r^2\vert \Psi(\vecr)\vert^2$. The spatial
correlation in the SCR is found to display a single peak for $\alpha =
0$, which clearly indicates a tightly bound state at the origin.  The
residual oscillations (for finite asymmetries) indicate that there is
no single bound state.  The increasingly oscillatory behavior of the
function $r^2\vert\Psi(\vecr)\vert^2$ in the ICR is consistent with
the notion of a transition from the BEC to the BCS regime. In the WCR,
strong oscillations are observed, reflecting the coherence of the BCS
state over large distances.

The coherence length, i.e., the spatial extension of a Cooper pair, is
then defined as
\be\label{eq:rms}
\xi_{\rm rms} = \sqrt{\langle
  r^2\rangle}.
\ee 
In the weak-coupling BCS regime the coherence length is given by the
well-known analytical formula
\bea 
\label{eq:xi_bcs}
\xi_a = \frac{\hbar^2
  k_F}{\pi m^* \Delta}.  
\eea
Table \ref{tab:1} lists the values of coherence length in the three 
coupling regimes, as obtained from Eq.~\eqref{eq:rms} and through the 
BCS formula \eqref{eq:xi_bcs}.  The corresponding values of the mean 
interparticle distance listed in the Table are determined from the density.
\begin{table*}
\begin{tabular}{cccccccccc}
\hline
 & log$_{10}(\rho/\rho_0)$ & $k_F$[fm$^{-1}$] &$T$ [MeV] 
& $d$ [fm] & $\xi_{\rm rms}$ [fm] & $\xi_{a}$ [fm] & $\lambda$ [fm] 
& $\kappa_{\rm rms}$ & $\kappa_{\rm rms}^{\rm LOFF}$ \\
\hline\hline
WCR & $-0.5$ &0.91& 0.5 & 1.68 & 3.17 & 1.41 & 18.5 & 5.8 &  0.8 \\
ICR & $-1.5$  & 0.42& 0.5 & 3.61 & 0.94 & 1.25 & 30.5 & 32.4 & --\\
SCR & $-2.5$ & 0.20 & 0.2 & 7.79 & 0.57 & 1.79 & 50.5 & 88.2 & --\\
\hline\\
\end{tabular}
\caption{
  For each of the three regimes of coupling strength, corresponding 
  values are presented for the density $\rho$ (in units of nuclear 
  saturation density $\rho_0 = 0.16$ fm$^{-3}$), Fermi momentum $k_F$, 
  temperature $T$, mean interparticle distance $d$, and coherence 
  parameters $\xi_{\rm rms}$ and $\xi_{a}$.   
  At $\alpha = 0$ the values of the pairing gap $\Delta$ in the
  three regimes WCR, ICR, and SCR are respectively 9.39, 4.50, and
  1.44 MeV, while the corresponding values of the effective mass 
  (in units of the bare mass) are 0.903, 0.989, and 0.999.  
  In the regime WCR, the LOFF phase is found in the vicinity of 
  asymmetry  $\alpha = 0.49$, at which $\Delta = 1.27$ MeV and 
  $Q = 0.4$ fm$^{-3}$. 
}
\label{tab:1}
\end{table*}

We now apply the macroscopic equations of electrodynamics of
superconductors. Since neutron-proton Cooper pairs carry a single unit of 
charge rather than two, the relations between various quantities differ 
accordingly from those for ordinary superconductors. To keep track of these 
differences we denote the charge and mass of a Cooper pair by $e_{X}$ 
and $m_{X}$.

The gauge invariant momentum associated with the Cooper wave function
is given by 
\be\label{1a}
\bm p = m_{X}\bm v = \hbar \bm\nabla\chi - \frac{e_{X}}{c}\bm A ,\\
\ee
where $\chi$ the phase of the wave function and $\bm A$
the vector potential. Taking the curl of equations (\ref{1a}) yields
\be\label{2a}
\curl \bm p = m_{X}\curl \bm v =  2\pi \hbar~ n_\phi \bm \nu- \frac{e_X}{c}\bm B, \\
\ee
where $n_\phi$ stands for the number density of vortices, 
$\bm\nu=(\curl \bm v)\, /\vert \curl \bm v \vert$ and $\bm B=\curl\bm A$. 
The Maxwell equation for the magnetic field is
\be\label{3}
\curl \bm B = \frac{4\pi }{c} e n_c \bm v,
\ee
where $n_c$ is number density of charge carriers. Taking the curl of
Eq.~\eqref{3}, we find the analogue of the London equation 
\bea\label{4c}
\curl\curl \bm B 
& =& \lambda^{-2}\left( \phi^* n_\phi \bm \nu- \bm B\right),
\eea
where
\be
\lambda = \left[\frac{4\pi e e_X}{m_Xc^2}  n_c 
 \right]^{1/2},\quad
\phi^* = \frac{2\pi \hbar c}{e_X} .
\ee
In the case of interest, we have $e_X = e$ and $m_X = 2m^*$, where $m^*$ is a
nucleon effective mass.  Therefore, 
\be\label{eq:lambda}
\lambda = \left[\frac{\pi e^2}{m^*c^2} (1-\alpha)
 \right]^{1/2},\quad
\phi^* = \frac{2\pi \hbar c}{e} = 2\phi_0 ,
\ee
where $n_c \equiv \rho_p = (1-\alpha) \rho/2$ is the density of charge
carriers.  It is seen from \eqref{eq:lambda} that the quantum vortices
present in the neutron-proton condensate carry twice the quantum of
magnetic flux.

The values of $\lambda$ and corresponding values of $\kappa$ are given
in Table \ref{tab:1} for representative points in the the coupling
regimes WCR, ICR, and SCR.  The gap value $\Delta = 1.27~{\rm MeV}$
found for the LOFF phase is much smaller than its corresponding value
in the BCS state.  Accordingly, to obtain a value for the coherence
length in the LOFF phase, we have rescaled the $\xi_{\rm rms}$ value
appearing in the table for the WCR case by the ratio of the pairing
gap in the BCS state to that in the LOFF state, which is $9.39/1.27$.
This yields $ \xi_{\rm rms}^{\rm LOFF}\simeq 23.4$ fm, which implies
$\kappa_{\rm rms}^{\rm LOFF} = 0.8$.  Since the GL parameter is very
close to the critical value $1/\sqrt{2}$ this particular phase is
close to the transition to a type-I superconducting state for values
of the gap $\Delta \le 1$ MeV.

\section{Conclusions and perspectives}
\label{sec:conclusions}
\vskip 0.3cm

Low-density nuclear and neutron matter feature a rich phase diagrams
in the presence of isospin asymmetry or spin polarization
respectively.  These phase diagrams may contain a number of phases: the
translationally and rotationally symmetric, but spin/isospin polarized
BCS phase, the BEC phase containing quasi-deuterons, the current-carrying 
LOFF phase, and the phase-separated phase. 

In this contribution we have concentrated on the electrodynamics of
these phases across the phase diagram derived in
Refs.~\cite{Stein:2012wd,Stein:2014nla}. We have evaluated the
penetration depth of the magnetic field of nucleonic matter in various
regions of the phase diagram and combined it with the microscopically
computed coherence length to obtain the Ginzburg-Landau parameter. We
find that in the entire region of the phase diagram considered here
(with few exceptions, see below), the condensed phases are type-II
superconductors which must form Abrikosov quantum vortices. However,
because the neutron-proton Cooper pairs carry only a unit of charge,
these vortices will carry twice the quantum of circulation
$\phi_0$. This is the main feature distinguishing the neutron-proton
condensate from ordinary ``condensed matter'' superconductors and the
proton superfluid in the core of a neutron star. In
addition, we find that the LOFF phase is at the point of transition 
between the type-I and type-II 
superconducting states for gap values of the order of MeV.  Because
sufficiently large asymmetries will suppress the gap to values
$\Delta\le 1$ MeV, it is clear that this phase may eventually turn into a
type-I superconductor. This argument applies to practically any phase
with asymmetry, as $\xi\to \infty$ when $\Delta(\alpha)\to 0$ and
therefore one finds $\kappa\to 0$. However, the relevant values of
$\Delta$ might be extremely small for other phases.

In this contribution we have followed an approach based on phenomenological 
Ginzburg-Landau theory.  It would be interesting and informative to extend 
our study of the electrodynamics of neutron-proton condensates to the 
microscopic level by evaluating the response functions of the condensates 
to electromagnetic fields (photon) in terms of the Green's functions 
introduced in Sec.~\ref{sec:review}.

\section*{Acknowledgments}
\vskip 0.3cm

We are grateful to Martin Stein and Xu-Guang Huang for collaboration on the
topics covered in Sec.~\ref{sec:review}.  AS is supported by the
Deutsche Forschungsgemeinschaft (Grant No. SE 1836/4-1). JWC
acknowledges research support from the McDonnell Center for the Space
Sciences and expresses his thanks to Professor Jos\'e Lu\'is da Silva
and his colleagues in the Centro de Investiga\c{c}\~{a}o em
Matem\'{a}tica e Aplica\c{c}\~{o}es during extended residence at the
University of Madeira.

\section*{References}
\vskip 0.3cm

\providecommand{\newblock}{}

\end{document}